\documentclass[journal=jacsat,manuscript=article,layout=twocolumn]{achemso}

\usepackage[version=3]{mhchem} 
\usepackage{amsmath,amsfonts}
\usepackage{booktabs}
\usepackage{siunitx}
\usepackage{multirow}

\usepackage{subcaption}

\author{Igor Evangelista}
\affiliation{Department of Materials Science and Engineering, University of Delaware, Newark, DE 19716 (USA)}
\email{igor@udel.edu}
\author{Abdul Saboor}
\affiliation{Department of Physics and Astronomy, University of Delaware, Newark, DE 19716 (USA)}
\email{asaboor@udel.edu}
\author{Muhammad Zubair}
\affiliation{Department of Physics and Astronomy, University of Delaware, Newark, DE 19716 (USA)}
\affiliation{Department of Materials Science and Engineering, University of Delaware, Newark, DE 19716 (USA)}
\author{Intuon Chatratin}
\author{Ruiqi Hu}
\author{Dai Q. Ho}
\affiliation{Department of Materials Science and Engineering, University of Delaware, Newark, DE 19716 (USA)}
\alsoaffiliation{Faculty of Natural Sciences, Quy Nhon University, Quy Nhon 55113, Vietnam}
\author{Shoaib Khalid}
\affiliation{Princeton Plasma Physics Laboratory, Princeton, NJ 08540 (USA)}
\author{Ioanna Fampiou}
\author{Anderson Janotti}
\affiliation{Department of Materials Science and Engineering, University of Delaware, Newark, DE 19716 (USA)}
\email{janotti@udel.edu}
\title{Effects of uniaxial strain on monolayer transition-metal dichalcogenides revisited}

\keywords{transition-metal dichalcogenides, uniaxial strain, valley drift, hybrid DFT+SOC, photoluminescence}
\begin{document}




\begin{abstract}
Using hybrid density functional calculations including spin-orbit coupling, we compute the strain evolution of the band structure of monolayer 1H-phase transition-metal dichalcogenides,  \(\ce{MX2}\) (M= Mo, W; X= S, Se, Te), emphasizing an accurate reproduction of the quasiparticle band gap (as opposed to the excitonic optical gap). We show that tensile uniaxial strain applied along either the armchair or zigzag directions leads to a pronounced reduction of the fundamental gap, with the conduction-band edge generally exhibiting the stronger strain response. Both the conduction-band electron valleys (CBM) and the valence-band hole valleys (VBM) remain degenerate under uniaxial strain, while simultaneously drifting away from the high-symmetry $K$ point under strain ("valley drift"), such that the band extrema occur at nearby off-symmetry wave vectors. A minimal tight-binding model rationalizes the valley drift and the unequal electron- and hole-valley drift rates in the presence of strain, leading to indirect band gaps. In particular, for \ce{MoS2} the indirectness increases with tensile strain, providing a natural explanation for the experimentally observed decrease in photoluminescence intensity under uniaxial deformation. These results provide quantitative guidance for tailoring band structures for optoelectronic and quantum-defect applications.
\end{abstract}

\section{Introduction}

\begin{figure}
    \centering
    \includegraphics[width=3.2 in]{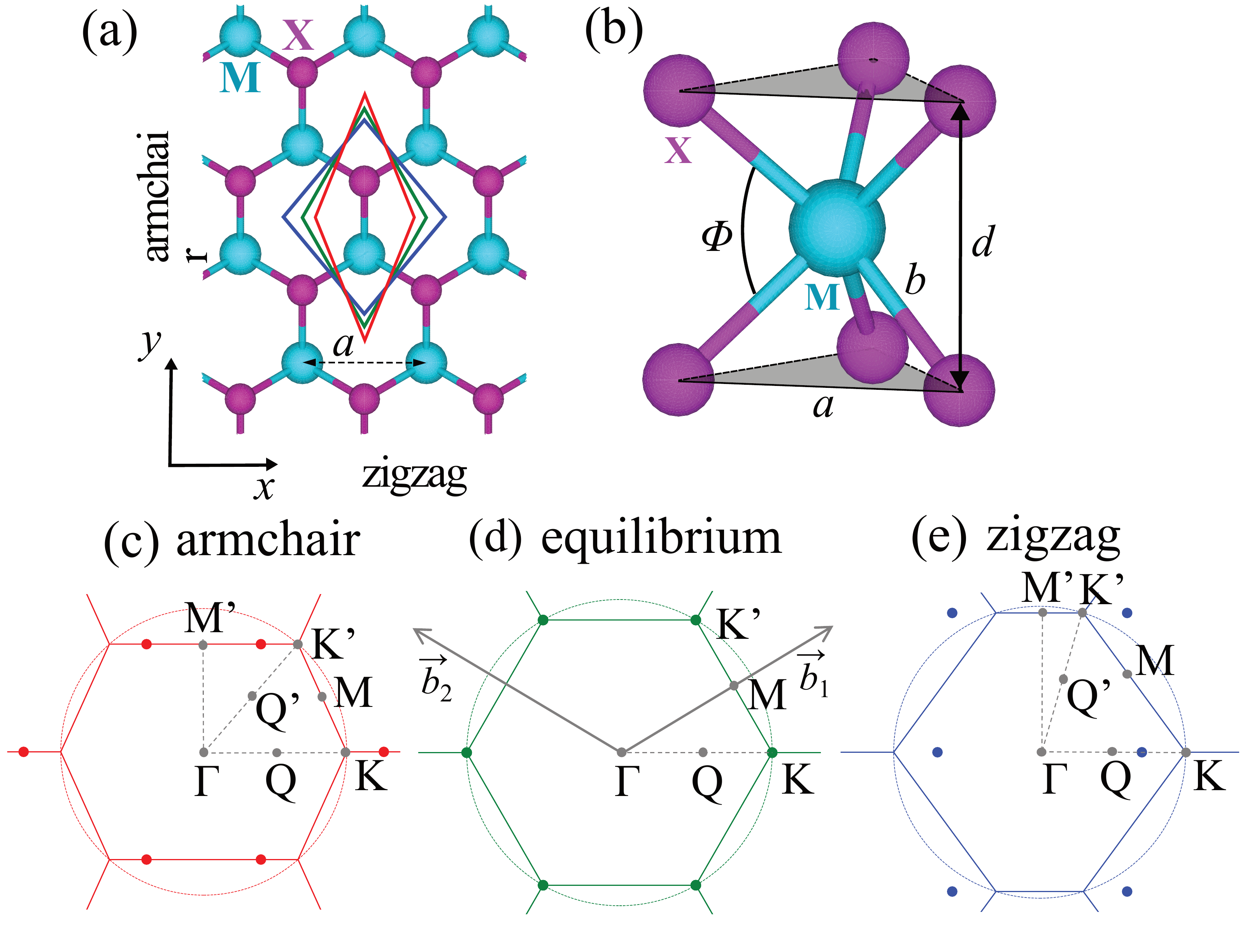}
    \caption{Crystal and reciprocal-space geometry of monolayer 1H-\ce{MX2}. (a) Top view and (b) side view of the trigonal-prismatic \ce{MX2} layer, with transition-metal atoms (blue) and chalcogen atoms (purple). The in-plane lattice vectors and structural descriptors used in this work (including the layer thickness $d$ and the chalcogen--metal--chalcogen bond angle $\phi$) are indicated in (b). (c-e) Brillouin zones (BZs) illustrating the strain-induced distortion of reciprocal space under tensile uniaxial loading: (d) unstrained (equilibrium) hexagonal BZ, compared to the distorted BZ for strain applied along (c) the armchair direction (red) and (e) the zigzag direction (green). The arrows $\mathbf{b}_1$ and $\mathbf{b}_2$ denote the reciprocal-lattice basis vectors. High-symmetry points ($\Gamma$, $K/K'$, $M/M'$, and the $Q/Q'$ points on $\Gamma$--$K$) are labeled for reference. The red dots in (c) and green dots in (e) schematically indicate the strain-induced drift of the electron and hole valleys away from the nominal $K$-corner, as analyzed in 
    Sec.~"Valley drift and strain-induced indirectness".}
    \label{fig1}
\end{figure}

Transition-metal dichalcogenides (TMDs) have emerged as a key platform in two-dimensional (2D) materials research because their layered crystal structure, strong spin-orbit coupling (SOC), and reduced dielectric screening enable electronic structures and optical responses that are qualitatively distinct from those of their bulk counterparts \cite{wang2012electronics, chhowalla2013chemistry}. Bulk semiconducting TMDs such as \ce{MoS2} crystallize in a van der Waals layered structure (space group $P6_3/mmc$, No.~194) in which each layer is a three-atom-thick X--M--X unit (X= chalcogen, M= transition metal), with the transition-metal plane sandwiched between two chalcogen planes and bound by strong intra-layer covalent/ionic bonding, while adjacent layers interact primarily through weak inter-layer dispersion forces. In the bulk, \ce{MoS2} is an indirect-gap semiconductor; upon thinning, the electronic structure evolves dramatically. In particular, monolayer \ce{MoS2} (space group $P\bar{6}m2$, No.~187), shown in Fig.~\ref{fig1}(a,b), becomes a direct-gap semiconductor with a (fundamental) quasiparticle gap of $\sim$2.4~eV \cite{huang2015bandgap,Note1}, making it attractive for nanoscale electronics and optoelectronics \cite{lebegue2009electronic, splendiani2010emerging, zhu2011giant}. This indirect-to-direct crossover, which persists down to the bilayer where the fundamental gap remains indirect, underlies the large enhancement of photoluminescence quantum efficiency in the monolayer limit \cite{pandey2020layer, mak2010atomically}.

Single- and few-layer TMDs are commonly obtained from bulk crystals by mechanical exfoliation or liquid-phase/solvent-assisted routes, in close analogy to graphene \cite{victor2023automated, li2013mechanical, li2014preparation, radisavljevic2011single, coleman2011two, lee2010frictional}. These fabrication pathways often introduce residual strain, either directly from the exfoliation/transfer steps or indirectly through adhesion, thermal-expansion mismatch, and interfacial interactions with the supporting substrate \cite{panasci_strain_2021, paradisanos_spatial_2016, castellanos2013local}. Because the band edges in TMDs derive from a strain-sensitive hybridization of metal $d$ and chalcogen $p$ orbitals, even moderate strain can substantially reshape the electronic and optical properties, including the magnitude and directness of the band gap, the effective masses and valley energetics, and therefore carrier mobility, conductivity, and photoluminescence efficiency \cite{yun2012thickness, peelaers2012effects, conley2013bandgap, he2013experimental}.

It has been demonstrated that monolayer TMDs can sustain remarkably large elastic deformations: freestanding membranes of \ce{MoS2} stretched with an atomic-force-microscope tip tolerate tensile strains approaching 11\% before failure \cite{bertolazzi2011stretching}, underscoring their high resistance to inelastic relaxation and fracture. Together with their direct band gap and exceptional in-plane stiffness/strength \cite{cooper2013nonlinear,li2013bonding}, this mechanical robustness has fueled strong interest in monolayer TMDs for flexible and deformable electronics and optoelectronics \cite{he2012fabrication, li2017graphene}. Beyond shifting band gaps, strain modifies photocarrier confinement landscapes \cite{chirolli2019strain}, enables the deterministic activation and tuning of single-photon emitters \cite{parto2021defect, peyskens2019integration, palacios2017large}, and has been implicated in emergent collective phenomena such as multi-valley superconductivity in gated or pressurized TMD systems \cite{pimenta2022electronic}.

Strain in monolayer TMDs can also be engineered intentionally, via external forces or stimuli, as a practical route to tailoring their electronic and optical properties \cite{roldan2015strain,rostami2015theory,He2016,blundo2021strain,Herrmann2025,conley2013bandgap}. Experimentally, strain has been introduced using a broad toolbox, including exfoliation or transfer onto patterned/topographically modulated substrates \cite{errando2021resonance}, piezoelectric actuation \cite{iff2019strain}, hydrostatic pressure in diamond-anvil cells \cite{pimenta2022electronic}, uniaxial elongation of elastomeric supports \cite{wang2015strain}, bending of flexible substrates \cite{conley2013bandgap}, and controlled wrinkling driven by thermal-expansion mismatch \cite{castellanos2013local}. Despite the qualitative clarity of the resulting trends, a quantitative determination of the applied strain field and, especially, of absolute band-edge shifts on a common energy scale remains challenging. In practice, strain is often inferred indirectly by comparing measured photoluminescence and/or infrared/Raman signatures to first-principles predictions for strain-dependent band gaps and phonon frequencies \cite{schmidt2016reversible, horzum2013phonon, yue2012mechanical, shi2013quasiparticle, amin2014strain, victor2025disentangling, conley2013bandgap}.

From a theoretical perspective, biaxial strain in monolayer TMDs is comparatively straightforward to treat, and the published trends are generally consistent and broadly available \cite{conley2013bandgap, schmidt2016reversible, horzum2013phonon, yue2012mechanical, shi2013quasiparticle, amin2014strain, he2013experimental, island2016precise}. In contrast, uniaxial loading, which necessarily generates coupled axial and transverse strains through the Poisson response, has been treated less uniformly. In some reports, the transverse response is effectively suppressed by adopting an unphysical zero-Poisson-ratio limit, while others do not consistently account for the distortion of the hexagonal Brillouin zone and the resulting relocation of the band extrema away from the high-symmetry $K$ point, leading to ambiguity in the identification of symmetry-inequivalent $k$-points under uniaxial stress \cite{wang2015strain, zhang2013giant,jiang2021uniaxial, shahriari2019investigation, lu2012strain, ghorbani2013strain, johari2012tuning, jena2019valleydrift}. In addition, conclusions that rely on semilocal functionals regarding a strain-driven crossover of the valence-band maximum from $K$ to $\Gamma$ under uniaxial strain should be treated with caution: the relative ordering of the $K$- and $\Gamma$-valence valleys is known to be sensitive to the exchange-correlation approximation and SOC, so an apparent $K$--$\Gamma$ crossing (and its "critical" strain) may reflect limitations of the functional rather than a robust physical trend \cite{lu2012strain, johari2012tuning, li2012ideal, jena2019valleydrift}. A further source of inconsistency is the frequent direct comparison between semilocal DFT band gaps and measured optical gaps, sometimes compounded by neglect of SOC \cite{lu2012strain, johari2012tuning, li2012ideal}. The optical gap differs from the fundamental quasiparticle gap by the exciton binding energy, which in monolayer TMDs is typically hundreds of meV \cite{mueller2018exciton, ugeda2014giant}; therefore, a quantitatively meaningful description of strain-induced gap changes and band-edge alignments requires quasiparticle-level gaps (or accurate surrogates) and SOC. These limitations in the existing literature on uniaxial stress provide the primary motivation for the present study.

Here, we use density functional theory (DFT) and hybrid DFT with spin-orbit coupling to systematically quantify the effects of uniaxial stress on the electronic structure of monolayer TMDs \ce{MX2} (M = Mo, W; X = S, Se, Te). We explicitly account for the strain-induced distortion of the hexagonal Brillouin zone and the resulting valley drift of both the conduction- and valence-band edges away from the high-symmetry $K$ point. Importantly, the electron (CBM) valleys remain degenerate in energy under uniaxial loading, and the same holds for the hole (VBM) valleys; however, the electron and hole extrema drift at different rates with strain. We attribute this asymmetry to the distinct orbital characters that dominate the band-edge states (and therefore to their different sensitivities to the strain-modified bonding and hybridization), which produces a growing momentum-space mismatch between the CBM and VBM and renders the fundamental gap increasingly indirect. Using the vacuum level as a common energy reference, we report absolute shifts of the CBM and VBM as functions of both axial and transverse strain for deformation along the zigzag and armchair directions, and we discuss the results in the context of available experimental data.

\section{Computational Approach}

Our first-principles calculations are based on density functional theory (DFT)\cite{hohenberg1964inhomogeneous,kohn1965selfconsistent}. Structural parameters (equilibrium lattice constants and internal coordinates) were obtained using the strongly constrained and appropriately normed (SCAN) meta-GGA exchange-correlation functional \cite{PhysRevLett.115.036402}. Electronic band structures were then computed with the Heyd-Scuseria--Ernzerhof screened hybrid functional (HSE) \cite{heyd2003hybrid,heyd2006hybrid} using a Hartree-Fock mixing parameter $\alpha=0.40$ and including spin-orbit coupling (HSE$\alpha$+SOC) in order to have a more accurate description of the quasiparticle band gaps and the strain-induced evolution of the band-edge valleys. The interaction between valence electrons and ionic cores was described within the projector augmented-wave (PAW) method \cite{blochl1994projector,kresse1999ultrasoft}, as implemented in the VASP code \cite{kresse1996efficient,Kresse1996_CMS6_15}.

Convergence of structural parameters and band gaps with respect to Brillouin-zone sampling and plane-wave cutoff was verified explicitly. We find that a $9\times 9\times 1$ $\Gamma$-centered $k$-point mesh and a plane-wave kinetic-energy cutoff of 600~eV yield converged results for all TMDs considered, both at equilibrium and under uniaxial stress. All structures were fully relaxed (cell shape consistent with the imposed strain state, and all ionic coordinates) until the total-energy change between electronic steps was below $10^{-7}$~eV and residual Hellmann-Feynman forces were below $10^{-4}$~eV\,\AA$^{-1}$. Hybrid-functional band structures were evaluated using the same $9\times 9\times 1$ mesh with SOC enabled; additional $k$-points along the chosen high-symmetry paths were included with zero weight to sample the band dispersion without affecting the self-consistent charge density.
 
The effect of uniaxial strain was modeled directly in the primitive hexagonal cell of the monolayer (Fig.~\ref{fig1}(a)), avoiding intermediate transformations to and from rectangular supercells. In the unstrained state, the in-plane primitive vectors are
$\vec{a}=\left(\frac{a}{2},\,\frac{\sqrt{3}a}{2}\right)$ and
$\vec{b}=\left(-\frac{a}{2},\,\frac{\sqrt{3}a}{2}\right)$, where $a$ is the equilibrium lattice parameter of \ce{MX2}. For uniaxial strain applied along the armchair direction ($y$), we impose a prescribed axial strain $\varepsilon_a$ by scaling the $y$ components of $\vec{a}$ and $\vec{b}$ accordingly, and then relax the transverse ($x$) components to determine the transverse strain $\varepsilon_t$ that minimizes the total energy. Conversely, for uniaxial strain applied along the zigzag direction ($x$), we fix the $x$ components of $\vec{a}$ and $\vec{b}$ to match the target $\varepsilon_a$ and relax the $y$ components to obtain the energy-minimizing $\varepsilon_t$. This procedure yields the coupled axial-transverse strain response (and hence the Poisson ratio) self-consistently for each $\varepsilon_a$, while preserving the correct hexagonal reciprocal-lattice distortion needed to construct the appropriate strained Brillouin zone. In particular, it eliminates the need for the rectangular-cell mapping used in previous works\cite{yue2012mechanical,jena2019valleydrift}, and reduces geometric bookkeeping errors when extracting the strain-dependent high-symmetry directions and valley locations.
To avoid spurious interactions between periodic images, a vacuum spacing of 20~\AA\ was included along the out-of-plane ($z$) direction. For each imposed axial strain (along zigzag or armchair), we determined the corresponding Poisson ratio and computed the SOC-included hybrid-functional band structure on the appropriately distorted Brillouin zone.

\begin{table*}[htb]
\centering
\begin{tabular}{|c|cc|cc|ccc|}
\hline
      & \multicolumn{2}{c|}{$a$ (\AA)} & \multicolumn{2}{c|}{atomic distances (\AA)} & \multicolumn{3}{c|}{$E_g$ (eV)}                                 \\ \hline
\ce{MoX2}  & \multicolumn{1}{c|}{calc.}  & exp.    & \multicolumn{1}{c|}{$d_{\rm X\!-\!X}$}   & $d_{\rm M\!-\!X}$  & \multicolumn{1}{c|}{calc.} & \multicolumn{1}{c|}{GW}   & exp.  \\ \hline
\ce{MoS2}  & \multicolumn{1}{c|}{3.174} & 3.2(1) \cite{huang2015bandgap} & \multicolumn{1}{c|}{3.102}  & 2.401 & \multicolumn{1}{c|}{2.28} & \multicolumn{1}{c|}{2.40} & 2.15(1) \cite{chiu2015determination} \\
\ce{MoSe2} & \multicolumn{1}{c|}{3.302} & 3.3(1) \cite{chen2017chemical}  & \multicolumn{1}{c|}{3.310}  & 2.525 & \multicolumn{1}{c|}{1.95} & \multicolumn{1}{c|}{2.08} & 2.18 \cite{ugeda2014giant}\\
\ce{MoTe2} & \multicolumn{1}{c|}{3.512} & 3.52 \cite{yu2017molecular}  & \multicolumn{1}{c|}{3.583}  & 2.706 & \multicolumn{1}{c|}{1.55} & \multicolumn{1}{c|}{1.65} & 1.72 \cite{Yang2015RobustEA} \\ \hline
\ce{WS2}   & \multicolumn{1}{c|}{3.161} & 3.153 \cite{gusakova2017electronic} & \multicolumn{1}{c|}{3.109}  & 2.398 & \multicolumn{1}{c|}{2.27} & \multicolumn{1}{c|}{2.46} & 2.38(6) \cite{hill2016band} \\
\ce{WSe2}  & \multicolumn{1}{c|}{3.289} & 3.29 \cite{huang2015large}  & \multicolumn{1}{c|}{3.311}  & 2.519 & \multicolumn{1}{c|}{1.90} & \multicolumn{1}{c|}{2.01} & 2.08(1)\cite{chiu2015determination} \\
\ce{WTe2}  & \multicolumn{1}{c|}{3.508} & --      & \multicolumn{1}{c|}{3.580}  & 2.703 & \multicolumn{1}{c|}{1.30} & \multicolumn{1}{c|}{1.45} & --    \\ \hline
\end{tabular}
\caption{Equilibrium lattice parameter $a$, layer thickness $d_{\rm X\!-\!X}$ (chalcogen--chalcogen distance across the metal plane), metal--chalcogen bond length $d_{\rm M\!-\!X}$, and band gap $E_g$ of monolayer 1H-\ce{MX2}. Structural parameters are obtained with SCAN, while band gaps are from HSE$\alpha$+SOC. Experimental data and representative quasiparticle $GW$ band gaps\cite{kim2021thickness} are included for comparison (no experimental structural/gap data are available for the metastable 1H-\ce{WTe2} phase).}
\label{tab1}
\end{table*}

\section{Results and Discussion}

We considered six prototypical group-VI transition-metal dichalcogenides in the semiconducting monolayer 1H-\ce{MX2} polymorph: \ce{MoS2}, \ce{MoSe2}, \ce{MoTe2}, \ce{WS2}, \ce{WSe2}, and \ce{WTe2}. For \ce{MoX2} (X= S, Se, Te) and for \ce{WX2} with X= S, Se, the 1H phase is the stable semiconducting form. In contrast, bulk \ce{WTe2} is known to crystallize in the metallic 1T$'$ orthorhombic structure (space group $P2_1/m$, No.~11) \cite{Brown1966_ActaCryst_WTe2_MoTe2,jana2015combined}. In the interest of a consistent, like-for-like comparison across the series, in particular, to isolate trends associated with chalcogen chemistry and spin-orbit coupling within a common trigonal-prismatic coordination environment, we nevertheless include \ce{WTe2} in the metastable 1H polymorph and report its computed properties in that structural setting.

\subsection{Unstrained monolayer TMDs}

The calculated (SCAN) equilibrium lattice parameters $a$, metal--chalcogen bond lengths $d_{\rm M\!-\!X}$, and layer thicknesses $d_{\rm X\!-\!X}$, together with the HSE$\alpha$+SOC band gaps $E_g$, are summarized in Table~\ref{tab1} and compared with available experimental measurements and published quasiparticle $GW$ results \cite{kim2021thickness}. Overall, the agreement is good, supporting the accuracy of our structural description and the quasiparticle character of the hybrid-functional gaps. Across a fixed metal series, the in-plane lattice parameter increases monotonically from S to Se to Te, whereas for a fixed chalcogen the Mo- and W-based compounds have very similar $a$ values (e.g., $a(\ce{MoX2})\approx a(\ce{WX2})$), indicating that the dominant structural control parameter within this family is the chalcogen size/chemistry. The small lattice mismatch between the common-chalcogen pairs \ce{MoX2} and \ce{WX2} implies that these combinations are particularly well suited for \ce{MoX2}/\ce{WX2} heterostructures, both for van der Waals (out-of-plane) stacking and for lateral (in-plane) junctions, where minimizing misfit strain and defect formation is essential \cite{tongay2014tuning}. The layer thickness $d_{\rm X\!-\!X}$ and the metal--chalcogen bond length $d_{\rm M\!-\!X}$ follow analogous trends.

\begin{figure}[ht!]
    \centering
    \includegraphics[width=3.4 in]{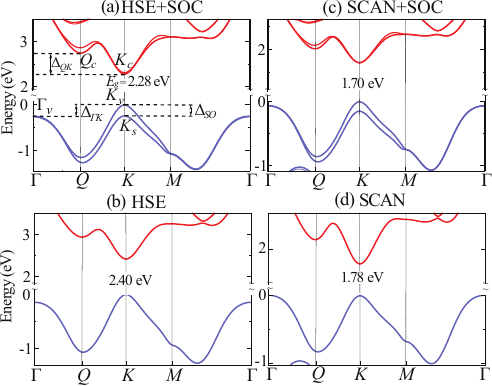}
    \caption{Electronic band structures of unstrained monolayer \ce{MoS2} calculated along $\Gamma$--$Q$--$K$--$M$--$\Gamma$ using different exchange-correlation functionals to highlight the roles of hybrid exchange and spin-orbit coupling (SOC): (a) HSE$\alpha$+SOC, (b) HSE$\alpha$ without SOC, (c) SCAN+SOC, and (d) SCAN without SOC. The valence-band maximum (VBM) is set to zero in all panels; red (blue) curves denote the lowest conduction (highest valence) bands. In (a), the direct gap at $K$ is labeled and the characteristic energy offsets $\Delta_{\Gamma K}$ and $\Delta_{QK}$, as well as the valence-band SOC splitting at $K$ ($\Delta_{\mathrm{SO}}$), are indicated; $K_v$ and $K_c$ mark the VBM and CBM at $K$, respectively.}
    \label{fig2}
\end{figure}

The band dispersion of unstrained monolayer \ce{MoS2} computed with HSE$\alpha$+SOC is shown in Fig.~\ref{fig2}(a); results obtained with SCAN (with and without SOC) are included for comparison. With HSE$\alpha$+SOC, SCAN+SOC, and HSE$\alpha$ (without SOC), the fundamental gap is direct at $K$, whereas SCAN without SOC predicts an indirect $\Gamma$--$K$ gap of 1.78~eV, in clear disagreement with experiment \cite{chiu2015determination}. This discrepancy reflects the decisive role of SOC in the relative alignment of the $K$ and $\Gamma$ valence valleys: SOC produces a substantial splitting at the VBM at $K$ ($\Delta_{\rm SO}\approx 250$~meV), which effectively raises the topmost $K$-valence state relative to the local maximum at $\Gamma$, thereby increasing $\Delta_{\Gamma K}$. Consistent with this picture, the separation between the direct $K$--$K$ gap and the indirect $\Gamma$--$K$ gap, $\Delta_{\Gamma K}$, is considerably larger in HSE$\alpha$+SOC than in SCAN+SOC (250~meV versus 65~meV). SOC also affects the conduction band, including the spin splitting at the secondary minimum at $Q$ (labeled $Q_c$ in Fig.~\ref{fig2}(a)). Because strain-induced direct-to-indirect transitions depend sensitively on the magnitude of $\Delta_{\Gamma K}$, approaches that neglect SOC or use functionals that severely underestimates the quasiparticle band gap tend to underestimate $\Delta_{\Gamma K}$ and consequently predict spurious or prematurely occurring direct-indirect crossovers at small strain \cite{yun2012thickness, johari2012tuning, li2012ideal,peto2019moderate, scalise2012strain,jena2019valleydrift}. Our calculated $\Delta_{\Gamma K}$ values for \ce{MoS2} (257~meV) and \ce{WSe2} (702~meV) are slightly higher than the corresponding experimental estimates (160~meV and 600~meV) \cite{chiu2015determination}; however, extracting the global VBM at $K$ and the local VBM at $\Gamma$ from scanning tunneling spectroscopy (STS) requires multiple tunneling conditions and carries a combined uncertainty of order 0.1~eV \cite{chiu2015determination}.

\begin{figure*}
    \centering
    \includegraphics[width=\textwidth]{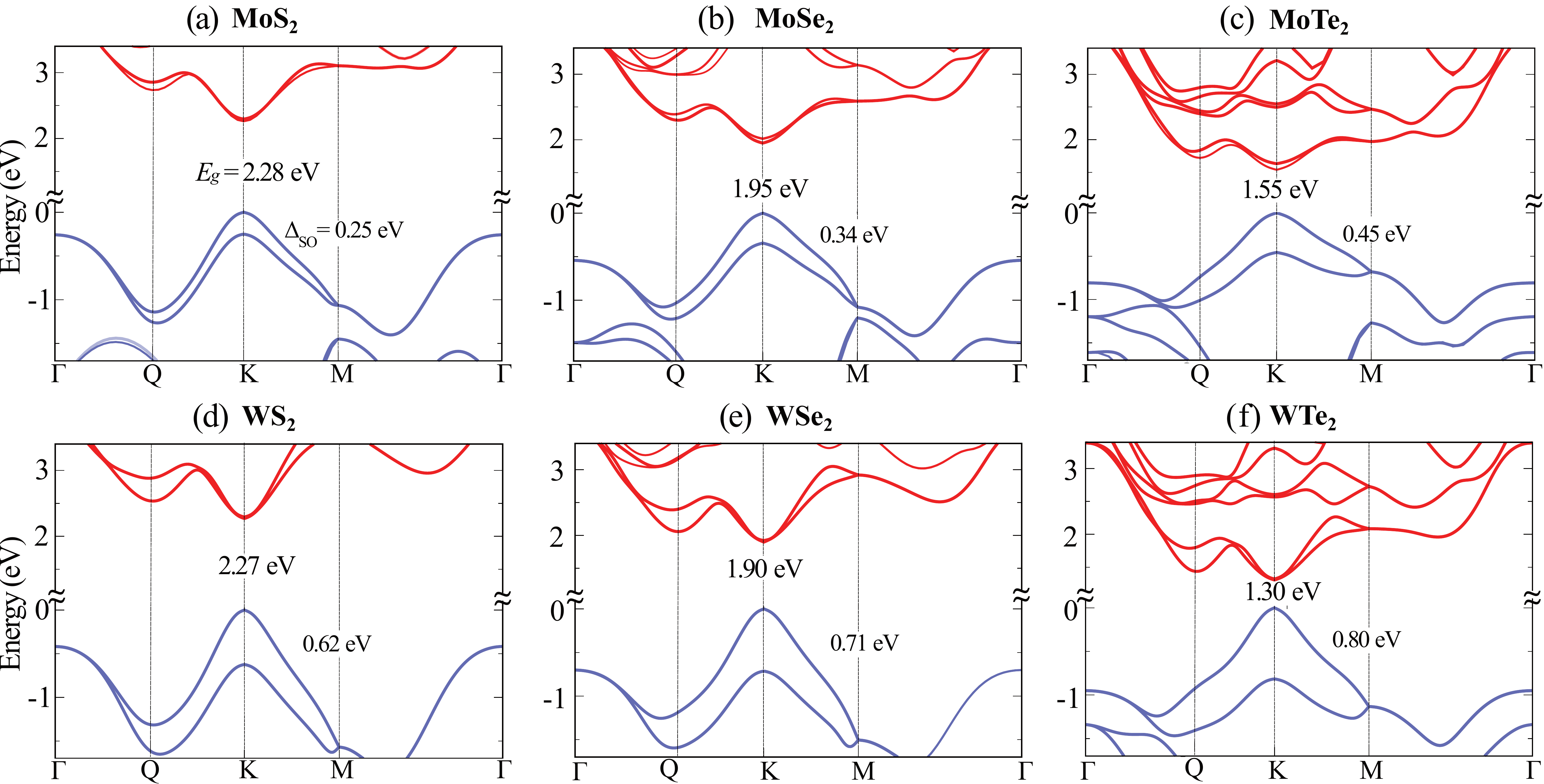}
    \caption{HSE$\alpha$+SOC band structures of unstrained monolayer 1H-\ce{MX2} for (a) \ce{MoS2}, (b) \ce{MoSe2}, (c) \ce{MoTe2}, (d) \ce{WS2}, (e) \ce{WSe2}, and (f) \ce{WTe2}, plotted along $\Gamma$--$Q$--$K$--$M$--$\Gamma$. Energies are referenced to the VBM (set to zero) in each panel. The fundamental quasiparticle gap $E_g$ and the valence-band SOC splitting at $K$ ($\Delta_{\mathrm{SO}}$) are annotated to facilitate comparison across the series, highlighting the systematic reduction of $E_g$ from S$\rightarrow$Se$\rightarrow$Te and the enhancement of SOC effects with heavier constituents.}
    \label{fig3}
\end{figure*}

Figure~\ref{fig3} compares the HSE$\alpha$+SOC band structures of the six monolayer 1H-\ce{MX2} compounds. The band gap decreases systematically from S to Se to Te, while for a fixed chalcogen the gaps of the Mo- and W-based materials are very similar (e.g., \ce{MoS2} versus \ce{WS2}). The valence-band spin-orbit splitting at $K$, $\Delta_{SO}$, increases markedly from S to Se to Te and also increases modestly when moving from Mo to W at fixed chalcogen, consistent with the stronger SOC in heavier elements. We further note that for \ce{WSe2} and \ce{WTe2} the lowest conduction-band minima at $K$ and $Q$ are nearly degenerate, in line with reports that monolayer \ce{WSe2} may be marginally indirect with the CBM at $Q$ \cite{zhang2015probing, hsu2017evidence, khalid2024deep}. Overall, the agreement with experiment and with quasiparticle $GW$ benchmarks (Table~\ref{tab1}) supports the reliability of the present approach for tracking strain-driven changes in valley energetics.

Inspecting the orbital character at high-symmetry points provides direct physical insight into both the band structure of the unstrained TMD and the response to strain, as discussed extensively in the literature \cite{Kang_2013,Padilha_2014,Kormanyos_2015,liu2015electronic}. The conduction-band minimum at $K$ ($K_c$) is dominated by the transition-metal $d_{z^2}$ orbital with smaller admixtures of chalcogen $p_x$ and $p_y$ character, whereas the valence-band maximum at $K$ ($K_v$) primarily derives from metal $d_{x^2-y^2}$/$d_{xy}$ orbitals with substantial hybridization with chalcogen $p_x$/$p_y$ states \cite{Padilha_2014, liu2015electronic}. Both $K_c$ and $K_v$ are predominantly antibonding \cite{Kang_2013, Padilha_2014}. By contrast, the top valence state at $\Gamma$ ($\Gamma_v$) originates mainly from metal $d_{z^2}$ and chalcogen $p_z$ orbitals and is bonding in the monolayer \cite{Padilha_2014}. In bilayer and thicker TMDs, interlayer coupling produces symmetric/antisymmetric combinations of these layer-resolved ($d_{z^2}$/$p_z$) states, and the higher-energy component can acquire antibonding character, contributing to the well-known thickness-driven indirectness of the band gap \cite{Padilha_2014}.

\subsection{Strained monolayer TMDs}

Uniaxial stress markedly reshapes the electronic structure of monolayer TMDs by perturbing the local bonding geometry and, consequently, the metal--chalcogen orbital hybridization that defines the band-edge valleys. By anisotropically modifying interatomic distances and bond angles, uniaxial loading alters the overlap among the transition-metal $d$ and chalcogen $p$ orbitals and shifts the relative energies (and, as discussed below, the $k$-space locations) of the key band-edge features at $K_c$, $K_v$, $\Gamma_v$, and $Q_c$. In this work, we focus on true uniaxial loading conditions: the monolayer is stretched along a single in-plane direction by an axial strain $\varepsilon_a$, while the lattice is fully relaxed in the perpendicular in-plane direction to determine the transverse strain $\varepsilon_t$. The associated Poisson ratio,
\begin{equation}
\nu=-\frac{\varepsilon_t}{\varepsilon_a},
\end{equation}
quantifies the lateral contraction/expansion induced by the applied axial strain and is positive for the systems studied here. In addition to in-plane relaxation, the monolayers respond by adjusting their out-of-plane thickness, which we characterize by the chalcogen--chalcogen separation $d_{\rm X\!-\!X}$ defined in Fig.~\ref{fig1}(b).

In the following, we restrict our attention to tensile axial strains up to 5\% and evaluate (i) the Poisson ratio, (ii) the evolution of the fundamental band gap, and (iii) the absolute shifts of the band-edge energies at $K_c$, $K_v$, $\Gamma_v$, and $Q_c$, referenced to the vacuum level. Uniaxial loading is applied along both the armchair and zigzag directions. 

We find that the Poisson ratio in monolayer TMDs depends only weakly on crystallographic orientation and, within the elastic range considered here, shows no appreciable dependence on strain. For \ce{MoS2}, for example, we obtain $\nu=0.22$ under armchair loading and $\nu=0.27$ under zigzag loading, consistent with prior theoretical studies \cite{yue2012mechanical,ccakir2014mechanical}. The slightly larger $\nu$ for zigzag strain implies a marginally stronger transverse contraction for a given axial stretch, indicating that \ce{MoS2} is modestly more compliant along the zigzag direction and correspondingly more rigid along the armchair direction. The same qualitative anisotropy is observed for the other TMDs considered here (Fig.~S1 of the Supplemental Material). These results contrast with previous calculations \cite{jena2019valleydrift}, which reported a strain-dependent Poisson ratio; such behavior would be unexpected for a material remaining in the linear elastic regime, where $\nu$ should be approximately constant.

\subsubsection{Basic electronic structure}

Uniaxial loading breaks the threefold rotational symmetry ($C_3$) of the 1H lattice and distorts the hexagonal Brillouin zone (BZ), as shown in Fig.~\ref{fig1}(c--e). This symmetry lowering has two practical consequences for band-structure analysis: (i) the \emph{positions} of the conventional high-symmetry points shift in reciprocal space as the reciprocal lattice vectors $\mathbf{b}_1$ and $\mathbf{b}_2$ deform, and (ii) points that were equivalent under $C_3$ rotations in the unstrained BZ split into distinct, strain-dependent sectors (most notably the $Q$ points on $\Gamma$--$K$). Importantly, however, the $K$ and $K'$ corners remain related by symmetry under uniaxial strain (see Fig.~\ref{fig1}(c--e)): time-reversal symmetry (and the remaining mirror symmetry for strain along armchair or zigzag) maps $K \leftrightarrow K'$, ensuring that the corresponding valleys remain degenerate in energy even though their locations (and the local dispersions) evolve with strain.

To capture these geometric and symmetry changes consistently, we compute strained band structures along the strain-adapted path $\Gamma \rightarrow Q \rightarrow Q' \rightarrow K \rightarrow K' \rightarrow M \rightarrow M'$, which resolves the strain-split valley directions on the distorted BZ. The relative separations among $K$- and $M$-type points depend on the loading axis: under armchair strain, the $K$ and $M$ corners move closer in reciprocal space, whereas under zigzag strain they separate further (Fig.~\ref{fig1}(c--e)). Unless stated otherwise, all strained dispersions reported below are obtained with HSE$\alpha$+SOC and are analyzed for tensile axial strains up to 5\%, with full transverse relaxation. While we compute strain responses across the full \ce{MX2} family, we focus the discussion of the electronic-structure consequences of \emph{uniaxial} strain on \ce{MoS2} as a representative example.

\begin{figure*}
    \centering
    \includegraphics[width=165mm]{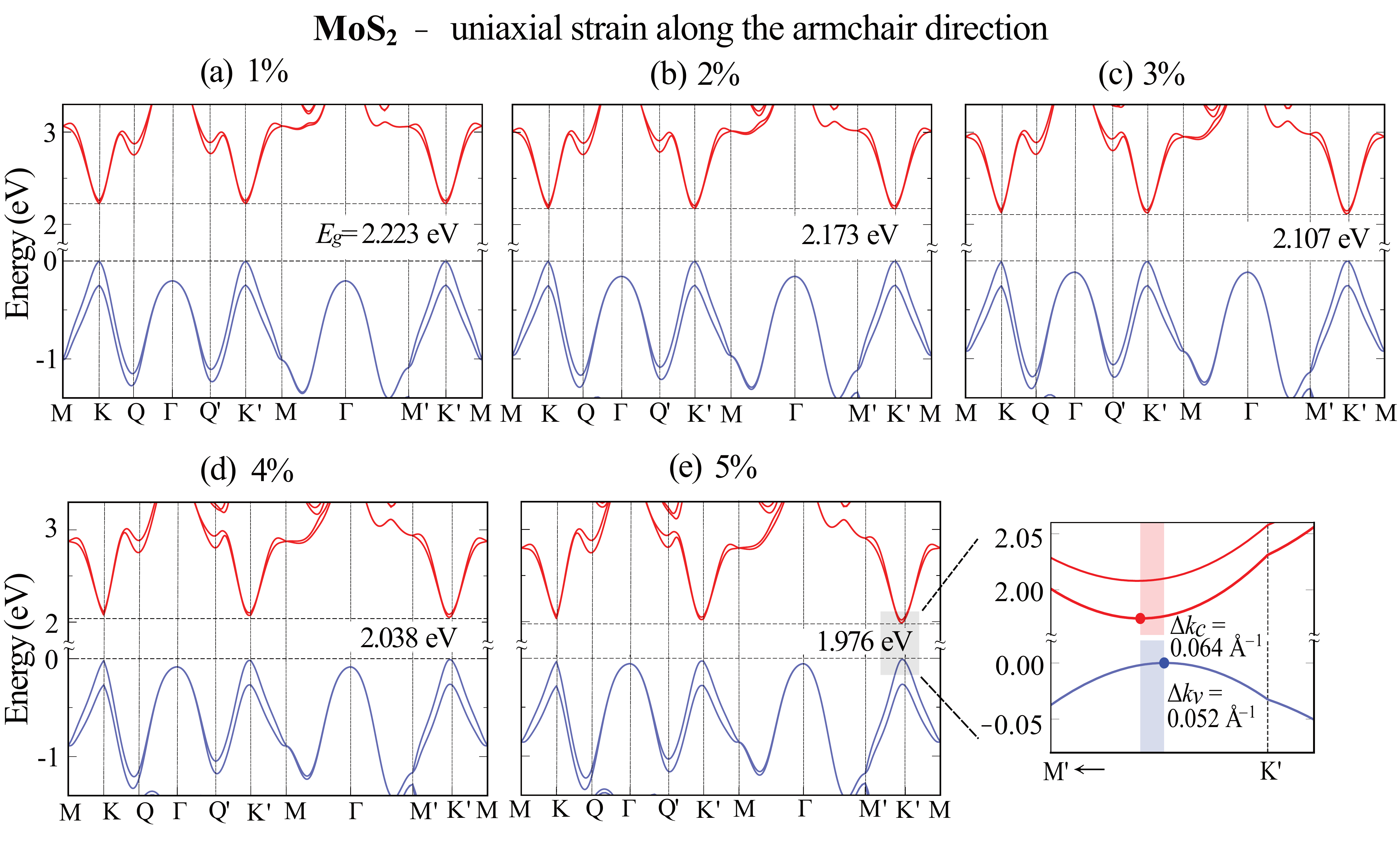}
    \caption{HSE$\alpha$+SOC band structures of monolayer \ce{MoS2} under tensile uniaxial strain applied along the armchair direction: (a) 1\% to (e) 5\%. Energies are referenced to the VBM (set to zero) in each panel, and the dispersions are plotted along the strain-adapted path $M\!\rightarrow\!K\!\rightarrow\!Q\!\rightarrow\!\Gamma\!\rightarrow\!Q'\!\rightarrow\!K'\!\rightarrow\!M'$.
    With increasing strain, the fundamental gap decreases, and the band-edge valleys exhibit a systematic \emph{valley drift} away from the nominal $K'$ point. The bottom-right inset (last panel) magnifies the vicinity of the strained valley extrema and explicitly marks the $k$-space displacement of the hole valley (VBM) and electron valley (CBM) from $K'$, quantified as $\Delta k_v$ and $\Delta k_c$, respectively (for 5\% strain: $\Delta k_v=0.052~\text{\AA}^{-1}$ and $\Delta k_c=0.064~\text{\AA}^{-1}$ along the indicated direction).}
    \label{fig4}
\end{figure*}

\begin{figure*}
    \centering
    \includegraphics[width=165mm]{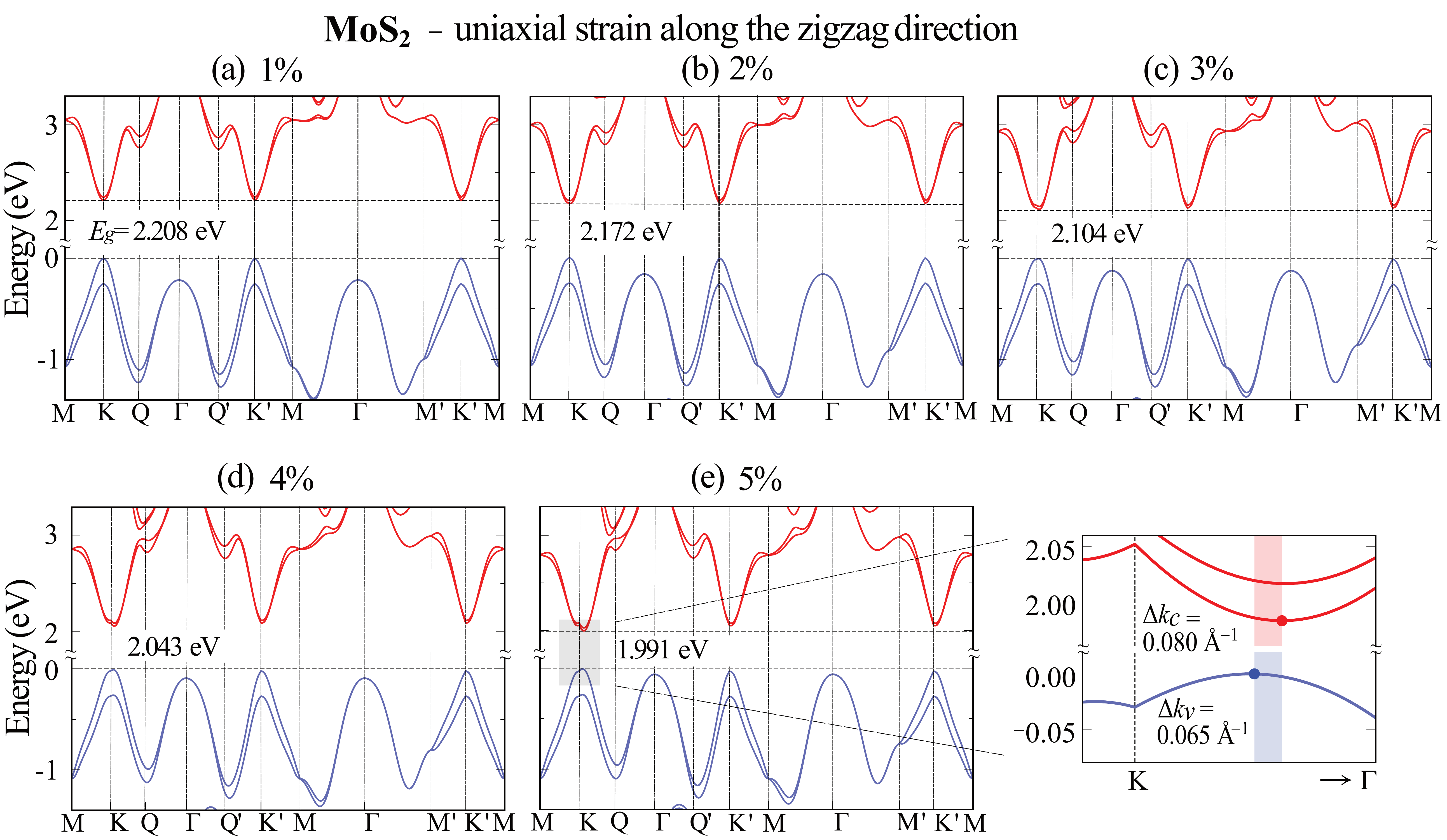}
    \caption{HSE$\alpha$+SOC band structures of monolayer \ce{MoS2} under tensile uniaxial strain applied along the \textbf{zigzag} direction: (a) 1\% to (e) 5\%. Energies are referenced to the VBM (set to zero) in each panel, and the dispersions are plotted along the strain-adapted path $M\!\rightarrow\!K\!\rightarrow\!Q\!\rightarrow\!\Gamma\!\rightarrow\!Q'\!\rightarrow\!K'\!\rightarrow\!M'$.
    The fundamental gap decreases with strain, and the band-edge valleys drift continuously away from the nominal $K$ point.
    The bottom-right inset (last panel) zooms into the vicinity of the valley extrema and highlights the strain-induced \textbf{hole-valley} (VBM) and \textbf{electron-valley} (CBM) displacements from $K$, labeled $\Delta k_v$ and $\Delta k_c$, respectively (for 5\% strain: $\Delta k_v=0.065~\text{\AA}^{-1}$ and $\Delta k_c=0.080~\text{\AA}^{-1}$ along the indicated direction), illustrating that the electron and hole valleys drift at different rates under uniaxial loading.}
    \label{fig5}
\end{figure*}

Figures~\ref{fig4} and \ref{fig5} summarize how the band edges of monolayer \ce{MoS2} evolve under armchair and zigzag tensile strain, respectively. In both loading geometries, the dominant trend is a rapid reduction of the fundamental gap with increasing strain, reflecting the sensitivity of the band-edge orbital hybridization to strain-modified bond lengths and angles. In the elastic strain window considered here (1--5\%), the band-edge valleys shift smoothly and approximately linearly away from the nominal $K$ points. At the smallest strains, the displacement is modest, and the extrema appear essentially at $K$ within the resolution of the plotted $k$-path (at $K'$ for armchair loading and $K$ for zigzag loading). As strain increases, the VBM and CBM move progressively off $K$ along strain-selected directions, e.g., toward the $M'$--$K'$ segment for armchair strain and toward the $K$--$Q$ segment for zigzag strain, so that the true extrema occur at nearby off-symmetry wave vectors. This nearly linear "valley drift" motivates locating band edges on the properly distorted Brillouin zone, rather than implicitly assuming that the extrema remain pinned to $K$ under uniaxial loading.

\subsubsection{Valley drift and strain-induced indirectness}\label{sec:valleydrift}

In the elastic uniaxial-strain regime (1--5\%), the band-edge valleys associated with $K/K'$ are displaced continuously from the nominal Brillouin-zone corners $K/K'$, while remaining degenerate in energy as required by time-reversal symmetry. This drift is clearly resolved in the strain-adapted band structures of \ce{MoS2} (Figs.~\ref{fig4}--\ref{fig5}) for both loading axes. In our lattice convention (Fig.~\ref{fig1}), "armchair" corresponds to axial deformation along $y$ and "zigzag" to axial deformation along $x$, with full transverse relaxation in each case.

Building on the strain-adapted Brillouin-zone construction described above, we now quantify the strain-induced displacement of the electron (CBM) and hole (VBM) valleys away from the nominal $K/K'$ points and examine how their unequal drift rates generate an increasing momentum mismatch, i.e., a progressively momentum-indirect fundamental gap. We note that the \emph{energies} of the symmetry-related valleys remain degenerate under uniaxial loading (i.e., the $K$-derived electron valleys remain equal in energy to each other, and likewise for the hole valleys), but their extrema are displaced in reciprocal space. This "valley drift" under uniaxial strain has also been reported previously\cite{zhang2013giant,jena2019valleydrift}, and it underlies strain-induced Berry-curvature polarization phenomena observed and modeled in monolayers \cite{son2019berrycurvature}. 

The microscopic origin of this displacement can be understood within a minimal tight-binding picture in which the $K$-valley band-edge states are \emph{phase-coherent} metal--chalcogen hybrids. Concretely, the band-edge Bl\"och eigenstate can be written as a superposition of Bl\"och sums of localized orbitals,
\begin{align}
\psi_{n\mathbf{k}}(\mathbf{r})
&=\sum_{\alpha} c^{(n)}_{\alpha}(\mathbf{k})\,\Phi_{\alpha\mathbf{k}}(\mathbf{r}),
\\[2pt]
\Phi_{\alpha\mathbf{k}}(\mathbf{r})
&=\frac{1}{\sqrt{N}}
\sum_{\mathbf{R}}
e^{i\mathbf{k}\cdot(\mathbf{R}+\boldsymbol{\tau}_\alpha)}
\,\phi_{\alpha}\!\left(\mathbf{r}-\mathbf{R}-\boldsymbol{\tau}_\alpha\right).
\end{align}
where $\alpha$ labels the relevant metal-$d$ and chalcogen-$p$ orbitals and the coefficients $c^{(n)}_{\alpha}(\mathbf{k})$ are, in general, complex. "Phase-coherent hybridization" means that the relative phases $\Delta\varphi_{\alpha\beta}(\mathbf{k})=\arg c_\alpha(\mathbf{k})-\arg c_\beta(\mathbf{k})$ are fixed by the eigenproblem and therefore allow amplitudes from different hopping pathways to interfere. This is made explicit in a minimal two-component basis (metal-like $|d\rangle$ and chalcogen-like $|p\rangle$), for which
\begin{align}
H(\mathbf{k})
&=
\begin{pmatrix}
\varepsilon_d(\mathbf{k}) & V(\mathbf{k})\\
V^*(\mathbf{k}) & \varepsilon_p(\mathbf{k})
\end{pmatrix},
\\[2pt]
V(\mathbf{k})
&=\sum_{j=1}^{3} t_j\, e^{i\mathbf{k}\cdot\boldsymbol{\delta}_j}.
\end{align}
with $\boldsymbol{\delta}_j$ the three nearest-neighbor bond vectors and $t_j$ the corresponding (strain-dependent) hopping amplitudes. The eigenvector relation implies
\begin{align}
\frac{c_p(\mathbf{k})}{c_d(\mathbf{k})}
&=\frac{E(\mathbf{k})-\varepsilon_d(\mathbf{k})}{V(\mathbf{k})},
\\[2pt]
\arg\!\left[\frac{c_p}{c_d}\right]
&=-\arg\!\left[V(\mathbf{k})\right],
\end{align}
i.e., the metal and chalcogen components are locked with a relative phase set by the $\mathbf{k}$-dependent hybridization (hopping) amplitude $V(\mathbf{k})$. In the unstrained lattice, $t_1=t_2=t_3\equiv t$ enforces at the BZ corners $K/K'$
\begin{equation}
V(\mathit{K}) = t\sum_{j=1}^{3} e^{i\mathit{K}\cdot\boldsymbol{\delta}_j}=0,
\end{equation}
because the three phases $\mathit{K}\!\cdot\!\boldsymbol{\delta}_j$ differ by $2\pi/3$ modulo $2\pi$, so that
$e^{i\mathit{K}\cdot\boldsymbol{\delta}_j}$ are separated by $120^\circ$ in the complex plane and add to zero (destructive interference among the three hopping channels). This cancellation pins the band-edge stationary point to the high-symmetry corner $\mathit{K}$.
Under uniaxial strain the anisotropic bonding makes $t_1,t_2,t_3$ inequivalent, the cancellation at $\mathit{K}$ becomes incomplete ($V(\mathit{K})\neq 0$), and a symmetry-allowed term linear in $\mathbf{q}=\mathbf{k}-\mathit{K}$ appears in the local dispersion, producing the observed valley drift. A convenient local form for the band-edge energy near $K$ is:
\begin{equation}
E(\mathit{K}+\mathbf{q}) \approx E_0 + \mathbf{v}\!\cdot\!\mathbf{q}
+\frac{1}{2}\sum_{i,j\in\{x,y\}} \left.\frac{\partial^2 E}{\partial q_i\,\partial q_j}\right|_{\mathit{K}} q_i q_j,
\label{eq:localdisp}
\end{equation}
where the curvature coefficients are evaluated at $\mathit{K}$ for the strained lattice, and
\begin{equation}
\mathbf{v}\equiv \left.\nabla_{\mathbf{k}} E(\mathbf{k})\right|_{\mathit{K}}
\end{equation}
is the strain-induced "drift velocity" (group-velocity term) that vanishes by symmetry in the unstrained ($C_3$) lattice but becomes finite when $t_1\neq t_2\neq t_3$. Minimizing Eq.~(\ref{eq:localdisp}) yields a displaced extremum at $\mathbf{k}=\mathit{K}+\delta\mathbf{k}$; in the present strain range, the resulting $\delta\mathbf{k}$ is approximately proportional to the applied uniaxial strain, consistent with the nearly linear drift observed in Figs.~\ref{fig4}--\ref{fig5}.

A key point for the fundamental-gap character is that the drift rates of the electron and hole valleys generally differ, because the relevant band-edge states have different orbital makeup (i.e., $d_{z^2}$,$p_x$/$p_y$ for electrons and $d_{x^2-y^2}$/$d_{xy}$,$p_x$/$p_y$ for holes) and therefore different sensitivities to the strain-modified hopping network. In \ce{MoS2}, for example, under armchair tensile strain the hole valley and electron valley both drift predominantly along the $K'$--$M'$ direction (along BZ edge), whereas under zigzag tensile strain both valleys drift along the $K$--$\Gamma$ direction (towards the center of the BZ), as shown schematically in Fig.~\ref{fig1}(c--e) and quantitatively in the inset of Figs.~\ref{fig4}--\ref{fig5}. More importantly, at 5\% armchair strain, we find the electron valley shifts by $|\delta k_e|=0.064~\text{\AA}^{-1}$ whereas the hole valley is displaced by $|\delta k_h|=0.052~\text{\AA}^{-1}$. In the case of zigzag strain, we find $|\delta k_e|=0.080~\text{\AA}^{-1}$ and $|\delta k_h|=0.065~\text{\AA}^{-1}$. 
Since the photon wavevector $k_\gamma=E/(\hbar c)$ is much smaller than the strain-induced $|\delta k_e-\delta k_h|$ already at a few percent strain, the lowest-energy excitons are pushed outside the radiative light cone, and the radiative recombination rate decreases continuously with increasing tensile strain.
Thus, even though each carrier type retains valley energy degeneracy, the \emph{relative} displacement of the CBM and VBM valleys converts the direct $K$--$K$ gap into a progressively indirect gap in momentum space as strain increases.

This strain-induced indirectness has immediate consequences for optical response. In the unstrained monolayer, interband transitions at $K$ lie inside the photon "light cone" (photon momentum is negligible on the Brillouin-zone scale), so radiative recombination is efficient and subject to the well-known valley-selective circular dichroism selection rules\cite{xiao2012coupled,mak2012control,zeng2012valley,palummo2015exciton}. When uniaxial strain pushes the band extrema away from $K$, the lowest-energy electron-hole recombination becomes momentum-indirect: direct photon emission from the true band extrema is suppressed unless assisted by phonons, disorder, or exciton center-of-mass momentum scattering. Equivalently, the fraction of the thermally populated exciton distribution that overlaps the radiative light cone decreases as $|\delta k_e-\delta k_h|$ grows, reducing the radiative rate and thereby diminishing photoluminescence intensity, a trend widely observed under tensile strain\cite{conley2013bandgap}. Moreover, because uniaxial strain reduces symmetry, it can also polarize Berry-curvature-related quantities in momentum space and modify valley-dependent optical and magneto-optical responses, as demonstrated via strain-tunable Berry curvature dipoles and valley magnetization in monolayer \ce{MoS2} \cite{son2019berrycurvature}. 

\subsubsection{Vacuum-referenced band-edge shifts across the \ce{MX2} family}

\begin{figure*}
    \centering
    \includegraphics[width=\textwidth]{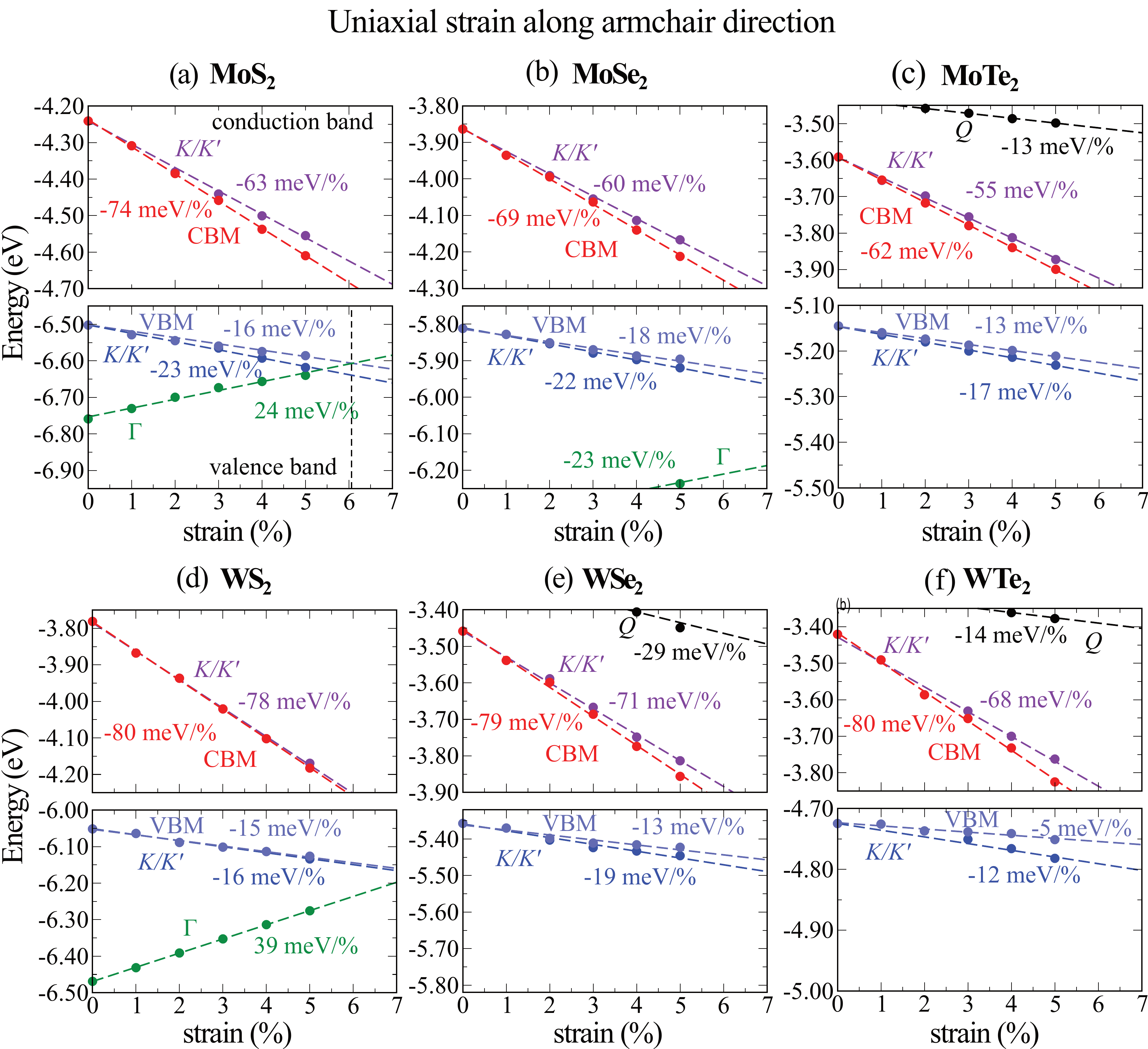}
    \caption{Vacuum-referenced evolution of the band edges of monolayer 1H-\ce{MX2} under 1--5\% tensile \emph{armchair} strain. Red symbols denote the CBM energy evaluated at the \emph{true} conduction-band minimum in the strained BZ (i.e., at the \emph{drifted} electron valley), while light-blue symbols denote the VBM energy evaluated at the \emph{true} valence-band maximum (the \emph{drifted} hole valley). Green symbols show the energy of the valence maximum at $\Gamma$ ($\Gamma_v$). For reference, the dashed purple/blue guides labeled $K/K'$ indicate the energies of the conduction- and valence-band edges \emph{at the nominal} $K/K'$ points; the growing separation between the drifted extrema (CBM/VBM) and their $K/K'$ values quantifies the increasing importance of valley drift with strain. The annotated slopes summarize the approximately linear band-edge response in the elastic strain window and provide strain coefficients for vacuum-level band alignment.}
    \label{fig6}
\end{figure*}

\begin{figure*}
    \centering
    \includegraphics[width=\textwidth]{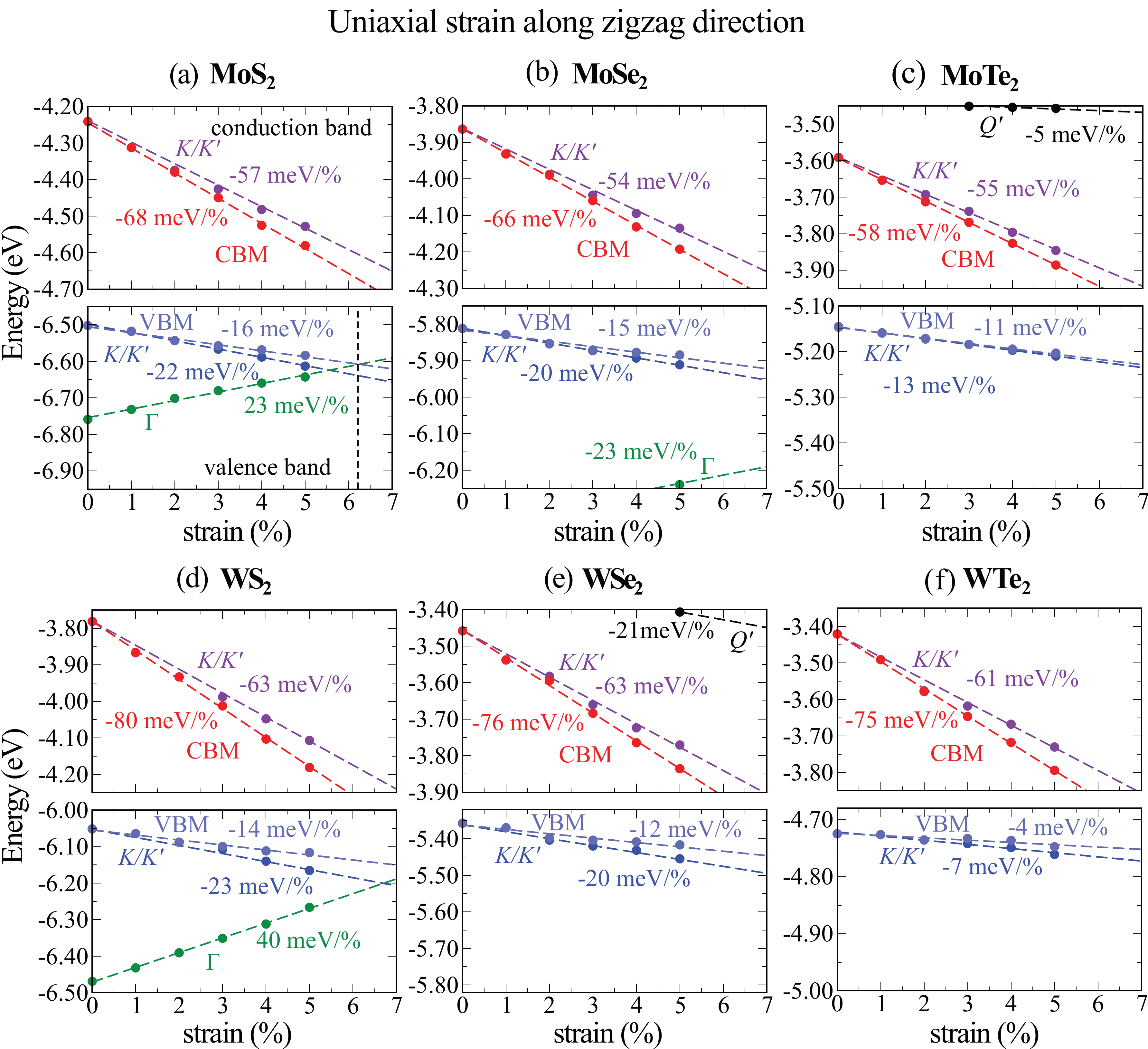}
    \caption{Vacuum-referenced evolution of the band edges of monolayer 1H-\ce{MX2} under 1--5\% tensile \emph{zigzag} strain. Red symbols denote the CBM energy at the \emph{drifted} conduction-band valley (true minimum), and light-blue symbols denote the VBM energy at the \emph{drifted} valence-band valley (true maximum). Green symbols show the $\Gamma$-valence maximum ($\Gamma_v$). Dashed guides labeled $K/K'$ indicate the corresponding band-edge energies \emph{evaluated at} the nominal $K/K'$ points for comparison. The increasing deviation between the drifted extrema and the $K/K'$-evaluated values with strain reflects the progressive relocation of the valleys in $k$-space under uniaxial loading. The annotated slopes summarize the nearly linear vacuum-referenced band-edge response within the elastic regime.}
    \label{fig7}
\end{figure*}

To place the strain response on an absolute energy scale and enable direct comparison to experiment, Figs.~\ref{fig6} and \ref{fig7} summarize the vacuum-referenced evolution of the key band energies evaluated at the nominal points ($K_c$, $K_v$, $\Gamma_v$, $Q_c$) together with the true (drifted) CBM and VBM that define the fundamental gap under uniaxial strain for all six monolayer 1H-\ce{MX2} compounds under 1-5\% tensile uniaxial strain applied along the armchair and zigzag directions, respectively. Reporting absolute band-edge positions, rather than only the fundamental gap, is critical for assessing strain-tunable band alignment (electron affinity and ionization energy) and for connecting to probes that reference the vacuum level, such as photoemission/inverse photoemission and scanning tunneling spectroscopy. (For context, we also consider biaxial tensile strain, which preserves the full hexagonal symmetry and therefore provides a useful baseline for disentangling symmetry-lowering effects unique to uniaxial deformation. The results are in Fig.~S2 of the Supplemental Material.)

Across the entire family, a robust, technologically relevant trend emerges: tensile strain shifts the conduction-band edge downward (i.e., toward a larger electron affinity) substantially more rapidly than it shifts the valence-band edge, so that the reduction in $E_g$ is primarily driven by the CBM response. The approximately linear slopes annotated in Figs.~\ref{fig6} and \ref{fig7} quantify this ``CBM-dominated'' gap renormalization in the elastic regime. In contrast, the valence-band response is more nuanced: the \emph{drifted} VBM exhibits comparatively small slopes, while the $\Gamma$-valence maximum ($\Gamma_v$) shifts in the opposite direction. This distinct behavior of $\Gamma_v$ is consistent with its different orbital character (predominantly metal $d_{z^2}$ and chalcogen $p_z$) and its enhanced sensitivity to strain-induced changes in out-of-plane geometry (layer thickness) and $p_z$-mediated hybridization. The $Q$-conduction valley ($Q_c$) also shows a material- and direction-dependent response, which is especially relevant for compounds where $K_c$ and $Q_c$ are nearly degenerate at equilibrium (e.g., \ce{WSe2}, \ce{WTe2}), as small strains can reorder the competing conduction minima.

It is also useful to distinguish two conceptually different routes to an "indirect" gap under tensile strain. In the widely cited interpretation of strained \ce{MoS2}, the apparent loss of photoluminescence is attributed to an energetic crossover in which the indirect $\Gamma$--$K$ transition becomes lower than the direct $K$--$K$ transition (i.e., $\Gamma_v$ rises relative to $K_v$ and/or the relative conduction valleys reorder), producing an \emph{energy-indirect} fundamental gap \cite{conley2013bandgap,jena2019valleydrift}. In our analysis, by explicitly tracking the \emph{drifted} band extrema on the distorted Brillouin zone, we find an additional and more continuous mechanism: even when the relevant band edges remain $K$-derived in energy, uniaxial strain displaces the CBM and VBM away from $K/K'$ with different drift rates, creating a growing electron-hole momentum mismatch $|\delta k_e-\delta k_h|$. This \emph{momentum-space} indirectness progressively pushes the lowest-energy electron-hole pairs outside the radiative light cone ($|\delta k_e-\delta k_h| \gg k_\gamma$), thereby reducing the fraction of excitons that can recombine radiatively. As a result, the valley-drift-induced momentum mismatch provides a natural explanation for the observed \emph{continuous} decrease of the photoluminescence intensity in monolayer \ce{MoS2} under tensile strain \cite{conley2013bandgap}, complementing (and in practice preceding) any eventual energetic crossover involving the $\Gamma$ valley \cite{jena2019valleydrift}.

Taken together, Figs.~\ref{fig6} and \ref{fig7} provide a compact "response map" for strain engineering: (i) the linear, vacuum-referenced slopes serve as transferable descriptors for predicting band offsets under moderate uniaxial deformation, and (ii) they establish the energetic backdrop against which the band-edge valleys simultaneously relocate in $k$-space. 

\section{Conclusion}

In summary, we revisited the electronic-structure response of monolayer 1H-phase TMDs to uniaxial strain using a methodology designed to avoid two common pitfalls: (i) improper treatment of the strain-induced Brillouin-zone distortion and the corresponding high-symmetry path, and (ii) band-gap and valley-energy inaccuracies associated with semilocal functionals and/or neglect of spin-orbit coupling. Using HSE$\alpha$+SOC for electronic structure, we obtain quasiparticle-level gaps and a consistent description of the strain-modified valley landscape.
For tensile uniaxial strains up to 5\% along either armchair or zigzag directions, we find that the fundamental gap decreases systematically across the \ce{MX2} family, driven predominantly by a strong downward shift of the CBM on an absolute (vacuum-referenced) energy scale, while the VBM shifts more weakly. The $\Gamma_v$ valley exhibits a qualitatively distinct response relative to the VBM near $K/K'$, reflecting its different ($d_{z^2}/p_z$-dominated) orbital character and sensitivity to out-of-plane structural relaxation. Beyond these energetic shifts, uniaxial strain induces a smooth, approximately linear drift of both electron and hole valleys away from the nominal $K/K'$ points on the distorted Brillouin zone, while preserving the energy degeneracy between time-reversal-related valleys. Because the electron and hole valleys drift at different rates, the lowest-energy transition becomes increasingly momentum-indirect, providing a natural mechanism for the experimentally observed reduction of photoluminescence intensity under tensile strain. 
Overall, the combination of (i) correct BZ geometry under uniaxial loading, (ii) vacuum-referenced band-edge shifts, and (iii) quasiparticle-level gaps with SOC yields quantitative guidance for strain-tunable band alignment and optical response in monolayer TMDs, and provides a reliable foundation for strain engineering in optoelectronics and quantum-defect applications.

\begin{acknowledgement}

This work was supported by the National Science Foundation award \#OIA-2217786, and the NSF UD-CHARM University of Delaware Materials Research Science and Engineering Center (MRSEC) Grant No. DMR2011824.  We also acknowledge the use of computational resources from the National Energy Research Scientific Computing Center (NERSC), a Department of Energy Office of Science User Facility, through the NERSC award BES-ERCAP 0034471 (m5002), and the DARWIN computing system at the University of Delaware, which is supported by the NSF Grant No.~1919839. S.K. acknowledges funding from the U.S. Department of Energy, Office of Science, Fusion Energy Sciences and Basic Energy Sciences, as part of the Extreme Lithography \& Materials Innovation Center (ELMIC), a Microelectronics Science Research Center (MSRC), under contract number No. DEAC02-09CH11466.

\end{acknowledgement}

\begin{suppinfo}

Supplemental material available.

\end{suppinfo}

\bibliography{UniaxialTMD}

\end{document}


\newpage
\begin{figure*}
    \centering
    \includegraphics[width=\textwidth]{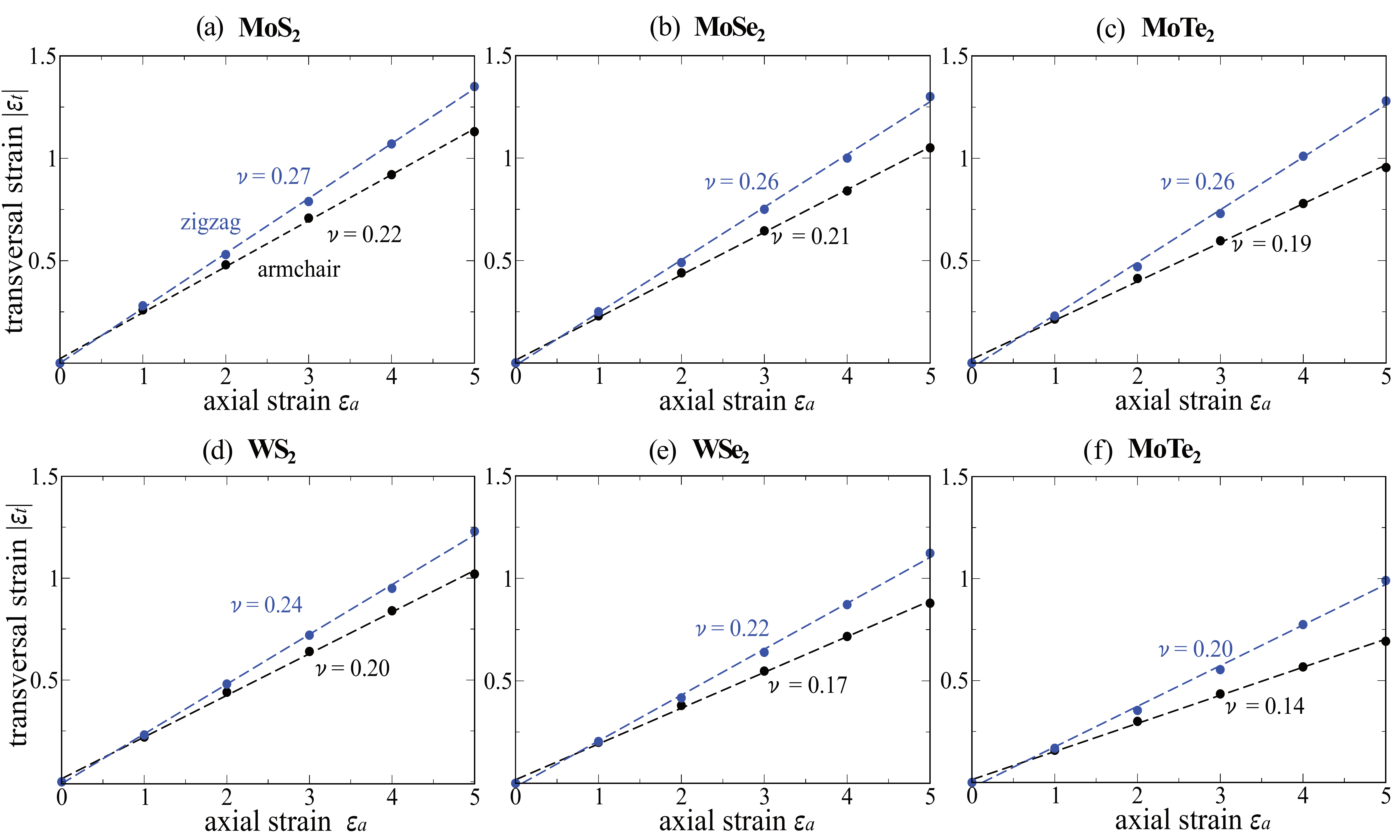}
    \caption{Transverse strain response and Poisson ratios of monolayer 1H-\ce{MX2} under uniaxial tensile loading (1--5\%). 
    Panels (a)-(f) show the transverse strain $\varepsilon_t$ as a function of the imposed axial strain $\varepsilon_a$ for (a) \ce{MoS2}, (b) \ce{MoSe2}, (c) \ce{MoTe2}, (d) \ce{WS2}, (e) \ce{WSe2}, and (f) \ce{WTe2}. 
    Black symbols/lines correspond to armchair loading and blue symbols/lines to zigzag loading (as defined by the strained primitive vectors used in the calculations). 
    Dashed lines are linear fits in the elastic regime; the slope yields the Poisson ratio $\nu=-\varepsilon_t/\varepsilon_a$ indicated in each panel, demonstrating a constant value and a weak dependence of $\nu$ on crystallographic orientation.}
    \label{figS1}
\end{figure*}

\begin{figure*}
    \centering
    \includegraphics[width=\textwidth]{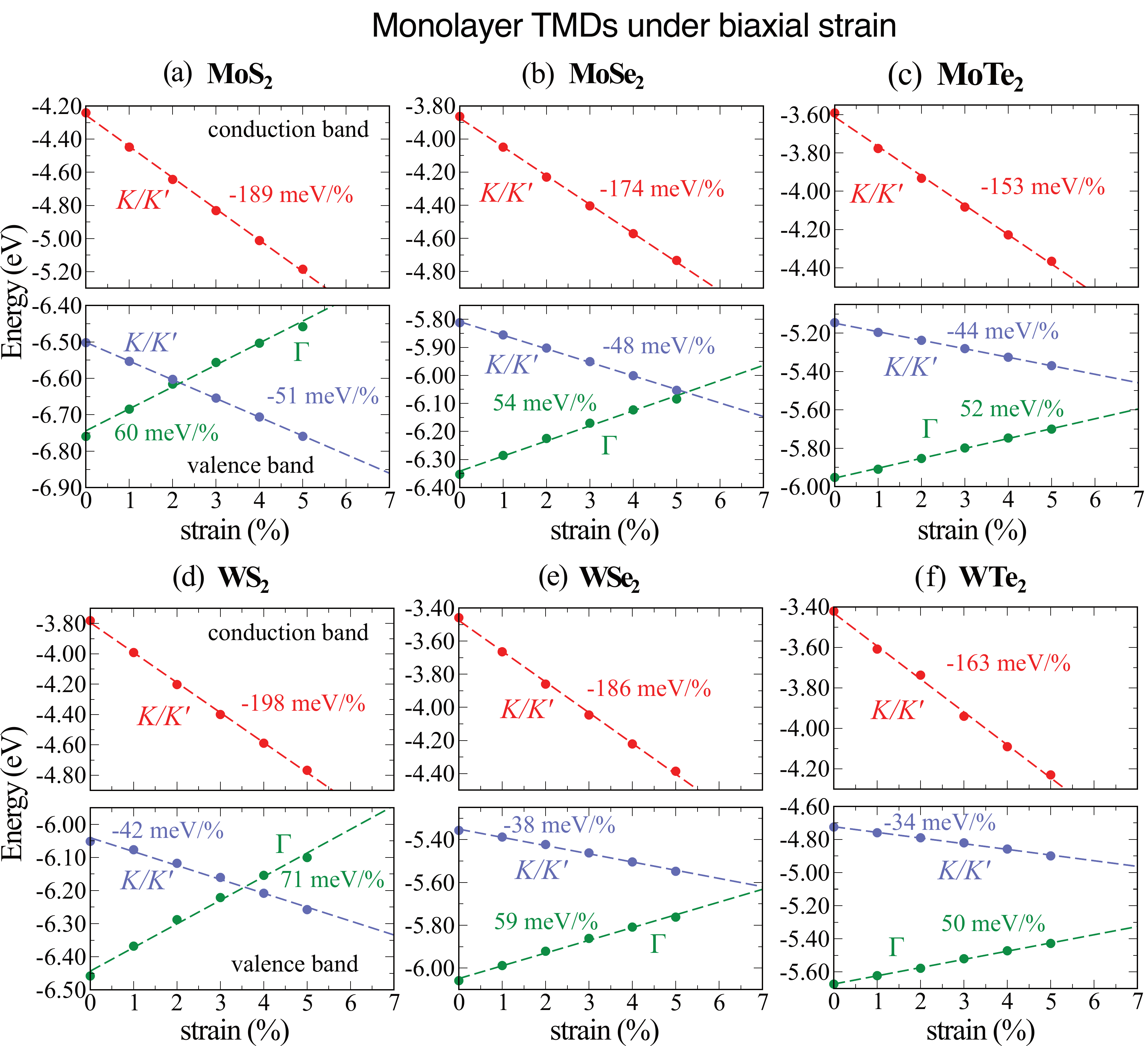}
    \caption{Vacuum-referenced band-edge shifts of monolayer 1H-\ce{MX2} under biaxial tensile strain (1--5\%). 
    Panels (a)--(f) correspond to \ce{MoS2}, \ce{MoSe2}, \ce{MoTe2}, \ce{WS2}, \ce{WSe2}, and \ce{WTe2}. 
    For each compound, the conduction-band minimum (CBM) at $K$ is shown in red, the valence-band maximum (VBM) at $K$ in blue, and the valence-band maximum at $\Gamma$ ($\Gamma_v$) in green, all referenced to the vacuum level (energies in eV). The annotated slopes (meV/\%) are obtained from linear fits over 1--5\% strain and quantify the band-edge deformation potentials at the labeled symmetry points. Compared with the uniaxial cases (Figs.~6--7 in the main text), biaxial strain produces a systematically larger magnitude of $\mathrm{d}E_g/\mathrm{d}\varepsilon$ (faster gap reduction) because it preserves hexagonal symmetry while uniformly modifying all three in-plane bond directions, leading to a stronger net renormalization of the band edges. Because biaxial strain does not induce valley drift, the extrema remain pinned at the nominal high-symmetry points; nevertheless, biaxial loading can drive an \emph{energetic} direct-to-indirect transition when $\Gamma_v$ crosses above $K_v$. Such a $\Gamma_v$--$K_v$ crossover (and the associated change from a direct $K$--$K$ gap to an indirect $\Gamma$--$K$ gap) occurs at comparatively small strain within the 1--5\% window for \ce{MoS2}, \ce{MoSe2}, \ce{WS2}, and \ce{WSe2}, whereas it is not observed in this strain range for \ce{MoTe2} and \ce{WTe2}.}
    \label{figS2}
\end{figure*}

\begin{table}[]
\centering
\begin{tabular}{|c|c|c|c|}
\hline
\begin{tabular}[c]{@{}c@{}}Band gap\\ Variation\end{tabular} & \begin{tabular}[c]{@{}c@{}}Armchair\\ (meV/\%)\end{tabular} & \begin{tabular}[c]{@{}c@{}}Zigzag\\ (meV/\%)\end{tabular} & \begin{tabular}[c]{@{}c@{}}Biaxial\\ (meV/\%)\end{tabular} \\ \hline
\ce{MoS2}                                                         & 58                                                          & 52                                                        & 138                                                        \\ \hline
\ce{MoSe2}                                                        & 51                                                          & 51                                                        & 126                                                        \\ \hline
\ce{MoTe2}                                                        & 49                                                          & 47                                                        & 109                                                        \\ \hline
\ce{WS2}                                                          & 65                                                          & 66                                                        & 156                                                        \\ \hline
\ce{WSe2}                                                         & 66                                                          & 64                                                        & 148                                                        \\ \hline
\ce{WTe2}                                                         & 75                                                          & 71                                                        & 129                                                        \\ \hline
\end{tabular}
\caption{Strain coefficients for the fundamental band gap of monolayer 1H-\ce{MX2} in the elastic regime. 
Listed are the linear rates of change of the fundamental gap, $|\mathrm{d}E_g/\mathrm{d}\varepsilon|$ (in meV/\%), extracted from HSE$\alpha$+SOC band structures for 1--5\% tensile loading under (i) uniaxial strain along the armchair direction (with full transverse relaxation), (ii) uniaxial strain along the zigzag direction (with full transverse relaxation), and (iii) biaxial tensile strain (preserving hexagonal symmetry). 
Positive values indicate a reduction of $E_g$ with increasing tensile strain.}
\label{variation}
\end{table}